\documentstyle[editedvolume, psfig]{crckapb}

\begin{opening}
\title{AXAF in Context: A Revolution}\\

\author{Martin Elvis}
\institute{Harvard-Smithsonian Center for Astrophysics\\
           60 Garden St., Cambridge MA, 02138, USA}

\end{opening}

\runningtitle{The AXAF Revolution}

\begin{document}


\section{AXAF and the Promise of X-ray Astrophysics}

AXAF has been a long time, 20 years, in coming.  This has made
AXAF so overfamiliar that it is hard to see how revolutionary it
still is. Some even describe AXAF as the end of the line in X-ray
technology; a never to be repeated venture to high resolution.
This view is wrong. To see why we need to see where X-ray
astrophysics is going.

It should not be controversial to say that X-ray Astrophysics
has barely begun. Today's X-ray satellites have scored many
successes; yet astronomers still extract very little of the
information carried to us by X-ray photons; the subarcsecond
spatial, and $R$=1000 spectral resolution {\em routinely}
available at longer wavelengths, is not even begun in X-ray
astronomy. AXAF is our first step into that world.

Eventually X-ray telescopes will be built that have all three
qualities needed to fully inhabit X-ray astrophysics: many
sq. meters of effective area; sub-arcsecond angular resolution;
and $R$=1000-10,000 spectral resolution. Then the riches of the
atomic transitions in the X-ray band spectrum can be exploited
for {\em all} clases of X-ray source (Elvis \& Fabbiano
1996). Getting there is the problem.

The next NASA Great Observatory, the {\em Advanced X-ray
Astrophysics Facility} takes the first step, combining two
qualities: sub-arcsecond imaging (0.5$^{\prime\prime}$ HPD) with
high spectral resolution ($R$=1000), albeit with a modest
increase in area ($\sim$0.1~m$^2$). So AXAF will show us where
X-ray astrophysics can go.  In this context we can recognize AXAF
for what it is - a revolution. This short paper tries to show
just how revolutionary.

\section{The Power of AXAF}

Why is AXAF `Advanced'?  The AXAF optics put AXAF in a separate
league from all other X-ray missions- past, present or planned:
75\% of power within 1~arcsec {\em diameter} up to 10~keV. (This
above spec. performance is thanks to Hughes-Danbury, Kodak, and
the AXAF Mirror Scientist, Leon van Speybroeck).  The AXAF {\em
beam area} (the important quantity) is $\frac{1}{100}$ that of
the ROSAT HRI, $\frac{1}{1000}$ of XMM and $\frac{1}{10,000}$ of
ASCA.  As a result AXAF has unprecedented sensitivity and
resolution,  angular and spectral.

AXAF has essentially zero background ( $\sim$1
count/megasec/sq.arcsec) and hence high sensitivity.  For point
sources observations up to 2 weeks long will be photon limited.
In 5~minutes AXAF will reach 10 times fainter than the ROSAT All
Sky Survey, and will return a $\frac{1}{2}^{\prime\prime}$
position. At this flux there are 4 million sources available. The
deepest, megasecond, AXAF surveys will reach
$f_x\sim5\times$10$^{-17}$~c.g.s., about $\frac{1}{20}$ of the
ROSAT limit.  This is uncharted territory.  Our only guide is the
ROSAT fluctuations analysis (Hasinger et al., 1993). This shows
that we can confidently expect several thousand sources per
square degree, i.e. about 1/sq.arcmin, or 250 per ACIS-I
field. But what are these objects?

Whenever one can image 100 times the detail of {\em any} previous
telescope extraordinary things will be found. The step from ROSAT
to AXAF is equal to that from ground-based astronomy to {\em
Hubble Space Telescope}. {\em Hubble} images often give me the
feeling of looking up the answer in the back of the book. These
images tell us that there is structure on every scale in
astrophysics.  Many of the same {\em Hubble} objects are also
bright X-ray sources. Surely they won't lose all structure when
we look with X-rays?

\begin{quote}
{\small
\begin{figure}[b]
\psfig{file=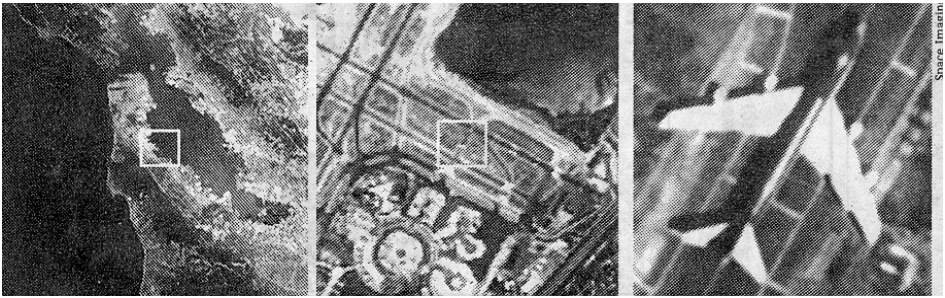,width=\textwidth,angle=-90}
\caption{San Francisco Bay, SFO airport, B747 at 30, 10,
1-meter resolution [N.Y.Times]}
\end{figure}
}
\end{quote}

\vspace{-8mm}

To get a more terrestrial perspective on what better angular
resolution can do, consider these images from spy satellites
published in the New York Times (February 10 1997, page 1.)  The
ratio of resolution in the three images correspond (from left to
right) to the ROSAT PSPC, the ROSAT HRI, and AXAF. While the
ROSAT HRI shows a peculiar and puzzling structure inside the
large feature found in the PSPC, time-resolved imaging at AXAF
resolution of the cross-shaped object in the lower panel makes
everything clear. {\em Hubble} shows that astrophysics works the
same way.

Choosing how to image a particular object with AXAF is complex.
AXAF carries three imaging instruments with multiple modes: 2
types of CCD (P.I. G. Garmire, Penn State) optimized for (1) low
energy (E$<$0.5~keV) response and high throughput (ACIS-S); (2)
good energy resolution ($\Delta E\sim$80~eV) and large
(16$^{\prime}\times$16$^{\prime}$) field-of-view (ACIS-I); and an
HRC microchannel plate (P.I. S. Murray , SAO) which is best for
fine detail ($<$0.5$^{\prime\prime}$), wide field
($\sim$25$^{\prime}$ dia.), and high time resolution (msec). The
ACIS (CCD) instrument gives $\sim$3-6 times, and the HRC
$\sim$twice, the PSPC or SIS count rate.

The AXAF transmission grating spectrometers give the first high
resolution data that other astronomers would call `spectra',
rather than broadband photometry, with useful area.  The AXAF
gratings have 100 times the spectral resolution of the ASCA SIS,
($R$=E/$\Delta E\sim 1000$) at 1~keV, and cover a seven times
broader energy range, 0.07-10~keV.  The line blending problems
that limit current spectra are largely gone at this resolution,
opening up fainter lines and so many physical diagnostics.  The
low energy (LETGS, P.I. B. Brinkmann, Utrecht) and high energy
(HETGS, P.I. C. Canizares, MIT) AXAF gratings have 20-200 times
greater area than their predecessors on {\em Einstein}.  At high
energies (E$>$0.5~keV) the HETGS gives count rates similar to the
ROSAT PSPC. The LETGS includes a region (0.07$<$E$<$0.5~keV) only
previously explored by the EUVE SWS spectrometer, and has an area
some 10 times larger.

AXAF is the first X-ray telescope with good simultaneous spatial
and spectral resolution.  Each ACIS CCD chip will have 250,000
independent beam areas. (ASCA has perhaps 16.) Not that AXAF has
the effective area to fill so many bins, but in a complex source
it will be possible to isolate structures, even the sinuous shock
fronts in supernova remnants and clusters of galaxies, and derive
their distinctive spectra.

The transmission gratings have spatial resolution too. They are
slitless spectrographs, familiar from optical prism surveys. The
image of the source is diffracted, so e.g. a supernova remnant
makes an image in each of the lines of its spectrum.  Complex
fields can provide enormous returns of data. A stellar cluster
can yield dozens of spectra. This is no simple analysis task -
the spectra overlap in space. Fortunately the CCD energy
resolution provides a third axis, and in this data cube the
spectra will almost all thread delicately past one another.

With AXAF's order of magnitude advances in angular resolution, in
spectral resolution and in both at once we can expect surprises.
Complex spectra will show up in unusual places; many spectra will
show features that are simply unknown, since laboratory work has
covered only a few of the transitions we will encounter with
AXAF; and many sources now thought of as simple will show complex
images, even whole new types of source. For example, {\em Hubble}
has shown that the bright stars in the Orion Trapezium are
surrounded by evaporating proto-planetary disks around nearby,
newly forming stars (Bally et al., 1997). Since the bright stars
are also bright X-ray sources, it is a simple prediction that
these evaporating proto-planetary disks will be shining in
fluorescent X-rays.

\section{AXAF is a Culmination, AND a Beginning}

Both technologically and scientifically AXAF is like {\em
Hubble}: both achieved 10 times improved resolution by using
heavy, rigid mirrors and were limited in area by the mirror
weight. But both demonstrate that high resolution is possible,
not end of the line. Scientifically, both let us see how complex,
yet comprehensible, the universe is.  There will be no going back
to less resolution, once we have seen {\em Hubble} and AXAF
images.

How do we get to more area with high resolution? Which axis
should we push on first? A first step is HTXS. With $\sim$3
sq.meters of collecting area it pushes the area dimension hard,
maintaining good spectral resolution. However angular resolution
will be limited. What is next?

The stumbling block is X-ray optics: we need 10~sq.meters of
effective area, yet must maintain arcsecond resolution (Elvis \&
Fabbiano 1996) and be light in weight.  Work on this challenging
goal is beginning in Europe, under the `XEUS' banner (M. Turner,
these proceedings).  Discussions in the NASA community are just
beginning. Certainly if we do not begin to develop the technology
for such a mission X-ray astronomy will wait another 20 years
before fulfilling the promise of the AXAF revolution.

\section{Your Turn}

After launch AXAF will quickly become a user driven observatory.
PV and Cal. observations will be public at once.  From month 5
onward 70\% of the time will be for Guest investigators,
increasing later to 85\%.  About 12~Msec of observing time is up
for bids in the first NASA announcement of opportunity, with a
deadline of 2 February 1998.  A Proposers Guide and an AXAF
simulator (`MARX') guide are available from the AXAF Science
Center ({\tt http://asc.harvard.edu}). AXAF is here at
last. Enjoy it.

\medskip
{\small
This work was supported in part by NASA contract NAS8-39073
(ASC).

\smallskip

\noindent
Bally J., et al. 1997,
{\tt http://www.cita.utoronto.ca/~johnston/orion.html}

\noindent
Elvis M., \& Fabbiano G., 1996, in {\em `Next Generation X-ray
Observatories'}, [U. Leicester], eds. M.J.L.Turner. \& M.G.Watson
XRA97/02, p. 33

\noindent
Hasinger G. et al., 1993, A\& A 288, 466
}

\end{document}